\documentclass[sigconf,9pt]{acmart}
\usepackage{graphicx} 
\usepackage{booktabs, multirow, array}
\usepackage{enumitem}
\usepackage{subcaption}
\AtBeginDocument{%
  \providecommand\BibTeX{{%
    \normalfont B\kern-0.5em{\scshape i\kern-0.25em b}\kern-0.8em\TeX}}}

\graphicspath{{./images/}} 
\newcommand{\sys}{\textsc{GUIDE}}

\definecolor{PrimaryBlue}{RGB}{13, 71, 161}
\definecolor{AccentOrange}{RGB}{230, 81, 0}
\definecolor{TealAccent}{RGB}{0, 105, 92}
\definecolor{PurpleAccent}{RGB}{74, 20, 140}
\definecolor{LightBlue}{RGB}{232, 244, 253}
\definecolor{LightOrange}{RGB}{255, 243, 224}
\definecolor{LightGray}{RGB}{248, 248, 248}
\definecolor{DarkGray}{RGB}{33, 33, 33}
\definecolor{MidGray}{RGB}{120, 120, 120}
\definecolor{GreenOK}{RGB}{27, 94, 32}
\definecolor{RedBad}{RGB}{183, 28, 28}
\definecolor{HeaderBg}{RGB}{13, 71, 161}

\copyrightyear{2026}
\acmYear{2026}
\setcopyright{cc}
\setcctype{by-nc-nd}
\acmConference[MobiSys Companion '26]{The 24th Annual International Conference on Mobile Systems, Applications and Services}{June 21--25, 2026}{Cambridge, United Kingdom}
\acmBooktitle{The 24th Annual International Conference on Mobile Systems, Applications and Services (MobiSys Companion '26), June 21--25, 2026, Cambridge, United Kingdom}
\acmDOI{10.1145/3812835.3814874}
\acmISBN{979-8-4007-2711-5/2026/06}

\begin{document}

\title{Poster: Practical Cross-Band Channel Prediction for AI-RAN via Physics-Guided Deep Unfolding}

\author{Ruiqi Kong, He Chen$^*$, Xiaojun Lin
}
\thanks{*Corresponding author.}
\affiliation{\institution{Department of Information Engineering, The Chinese University of Hong Kong, Hong Kong SAR, China}
    \country{Email:~\{rqkong, he.chen, xjlin\}@ie.cuhk.edu.hk}
}

\begin{abstract}
To make cross-band channel prediction practical for AI-native RAN, algorithms must generalize across diverse environments and support real-time inference. Existing approaches achieve one but not both. 
To bridge this gap, we introduce GUIDE, a physics-guided deep unfolding framework that embeds wireless channel physics into differentiable layers. Without retraining in unseen environments, \sys{} achieves $2.75\times$ beamforming gain than the deep learning-based baseline FIRE with only a slight increase in inference time, and $1.39\times$ beamforming gain than the strongest model-based baseline R2F2 while running over $1610\times$ faster.

\end{abstract}


  
\maketitle
\vspace{-0.5em}
\section{Introduction}


\begin{table}
	\centering
	\caption{Comparison of channel prediction methods.}
	\label{tab:comparison}
	\renewcommand{\arraystretch}{1.2}
	\vspace{-1em}
	\small
	\begin{tabular}{lcccc}
		\toprule
		\textbf{Property}
		& \textbf{R2F2}
		& \textbf{FIRE}
		& \textbf{\small HORCRUX}
		& \textbf{\sys{}} \\
		\midrule
		Env.\ generalization
		& \textcolor{GreenOK}{\checkmark}
		& \textcolor{RedBad}{$\times$}
		& \textcolor{GreenOK}{\checkmark}
		& \textcolor{GreenOK}{\checkmark} \\
		Fast inference
		& \textcolor{RedBad}{$\times$}
		& \textcolor{GreenOK}{\checkmark}
		& \textcolor{RedBad}{$\times$}
		& \textcolor{GreenOK}{\checkmark} \\
		Multipath-rich envs
		& \textcolor{RedBad}{$\times$}
		& \textcolor{RedBad}{$\times$}
		& \textcolor{RedBad}{$\times$}
		& \textcolor{GreenOK}{\checkmark} \\
		\bottomrule
	\end{tabular}
	\vspace{-1em}
\end{table}

Among the AI-native capabilities envisioned for AI-RAN, cross-band channel prediction is particularly valuable for frequency division duplexing (FDD) cellular systems~\cite{horcrux2024, fire2021}, where downlink (DL) channel state information (CSI) acquisition typically relies on explicit user feedback~\cite{optML2019}. In massive MIMO, this feedback incurs substantial bandwidth overhead and latency~\cite{horcrux2024}. The key physical insight enabling feedback-free FDD is that the underlying multipath parameters are \textit{frequency-independent}.
If these parameters can be accurately estimated from the uplink (UL) channel, the DL channel at any frequency can be reconstructed directly. 

Existing approaches adopt two broad paradigms, each with distinct trade-offs, as summarised in Table~\ref{tab:comparison}. \textit{Model-based methods} such as \textsc{R2F2}~\cite{r2f2_2016} and \textsc{Horcrux}~\cite{horcrux2024} explicitly exploit the physical structure: \textsc{R2F2} iteratively estimates multipath parameters, offering strong interpretability and cross-environment transferability, yet its 
greedy search is computationally prohibitive and collapses under dense multipath
; \textsc{Horcrux} improves robustness through sub-band neural estimators but retains a disjoint two-stage pipeline that 
is not jointly optimised, limiting accuracy and increasing latency. \textit{Black-box deep learning methods} such as \textsc{Fire}~\cite{fire2021} sidestep explicit modelling entirely, achieving fast inference by learning a direct 
UL-to-DL mapping, but at the cost of poor generalization; the learned mapping must be retrained whenever environments or system setups change.

This motivates a \textit{model-driven deep learning} approach that combines the physical interpretability and transferability of model-based methods with the speed and end-to-end trainability of data-driven ones. \sys{} achieves this via \textit{deep unfolding}: the iterative estimation of multipath parameters is mapped to multiple differentiable network layers with physics constraints embedded structurally. This makes \sys{} accurate, fast, and transferable to unseen environments without retraining the neural network itself. 

\vspace{-0.5em}
\section{\sys{} Design}

\begin{figure}[t]
	\centering
	\includegraphics[width=0.8\linewidth]{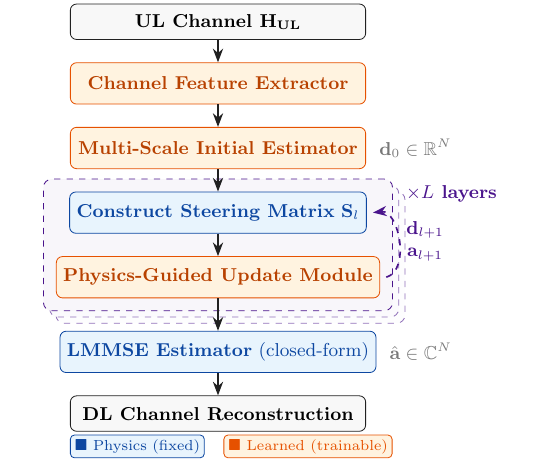}
	\vspace{-1em}
	\caption{
		\textbf{\sys{} architecture.}
		Blue blocks are physics-derived (differentiable but
		fixed-form); orange blocks are learned neural modules.
		The dashed purple box denotes the $L$-layer unfolding loop.
	}
	\label{fig:architecture}
	\vspace{-1em}
\end{figure}

\sys{} unfolds the iterative estimation of multipath parameters $\{ \text{distance } d_n, \text{complex attenuation } a_n\}$ into a stack of $L$ differentiable physics-guided layers trained end-to-end via backpropagation. Realizing this idea, however, raises two non-trivial challenges: \textit{(i)} determining which physical constraints to be embedded rigidly versus left to the data, and \textit{(ii)} ensuring stable gradient flow through layers after 
physical constraints embedding. The design of \sys{} is shaped by both, as shown in Fig.~\ref{fig:architecture}. 

The observed UL CSI matrix $\mathbf{H}_{\text{UL}}$ is first processed by a multi-scale CNN, capturing both local and global structure across subcarriers to produce a shared feature representation for all subsequent steps. Reliable initialization of multipath parameters is essential before unfolding: a poor starting point risks slow convergence or entrapment in local optima. A learned initializer layer therefore uses cross-attention to aggregate per-antenna features into $N$ path-specific estimates $\mathbf{d}_0$, placing the solution close to the true parameter manifold.

The unfolding stage then refines these estimates over $L$ iterations, directly confronting challenge~\textit{(i)}: embedding too many constraints reduces model capacity under environment mismatch, while embedding too few sacrifices cross-environment transferability. \sys{} resolves this tension by hard-coding only the steering matrix construction, encoding propagation distance via complex exponentials, while delegating residual correction to a lightweight learned 
module. Specifically, at each iteration $l$, the steering matrix $\mathbf{S}_l$ is constructed analytically from current estimates $\mathbf{d}_l$, and the residual $\mathbf{r}_l = \mathbf{H}_{\text{UL}} - \mathbf{S}_l \mathbf{a}_l$ provides a physically interpretable error signal. The learned corrector then outputs refined estimates $\mathbf{d}_{l+1}$, absorbing model mismatch without re-learning the underlying wave physics. Challenge~\textit{(ii)} is addressed by the analytical differentiability of the steering matrix: since each entry of $\mathbf{S}_l$ is a composition of smooth complex exponentials, gradients propagate cleanly through all $L$ physics layers during backpropagation without approximation. Once geometric parameters converge, the complex path attenuations $\hat{\mathbf{a}}$ are recovered in closed form via the LMMSE estimator, which remains fully differentiable throughout training.

\section{Experimental Evaluation}


We synthesize 80K training channels under the CDL-B 
non-line-of-sight (NLoS) model with 10\,MHz bandwidth, 26 subcarriers, 
and SNR $\in [10, 20]$\,dB. We evaluate generalisation on over-the-air 
measurements collected in a complex office environment under NLoS conditions 
using a Xilinx ZC706 SDR with four antennas. Uplink channels are measured at 2.412\,GHz and downlink channels are predicted at 2.422\,GHz. 
No retraining is performed on real-world data. All training and inference are conducted on an NVIDIA RTX 4070 GPU. Three baselines are evaluated: R2F2~\cite{r2f2_2016}, FIRE~\cite{fire2021}, and HORCRUX~\cite{horcrux2024}, with beamforming gain as the metric.

\begin{figure}[t]
	\centering
	\subfloat[Beamforming Gain.]{\includegraphics[width=0.5\linewidth]{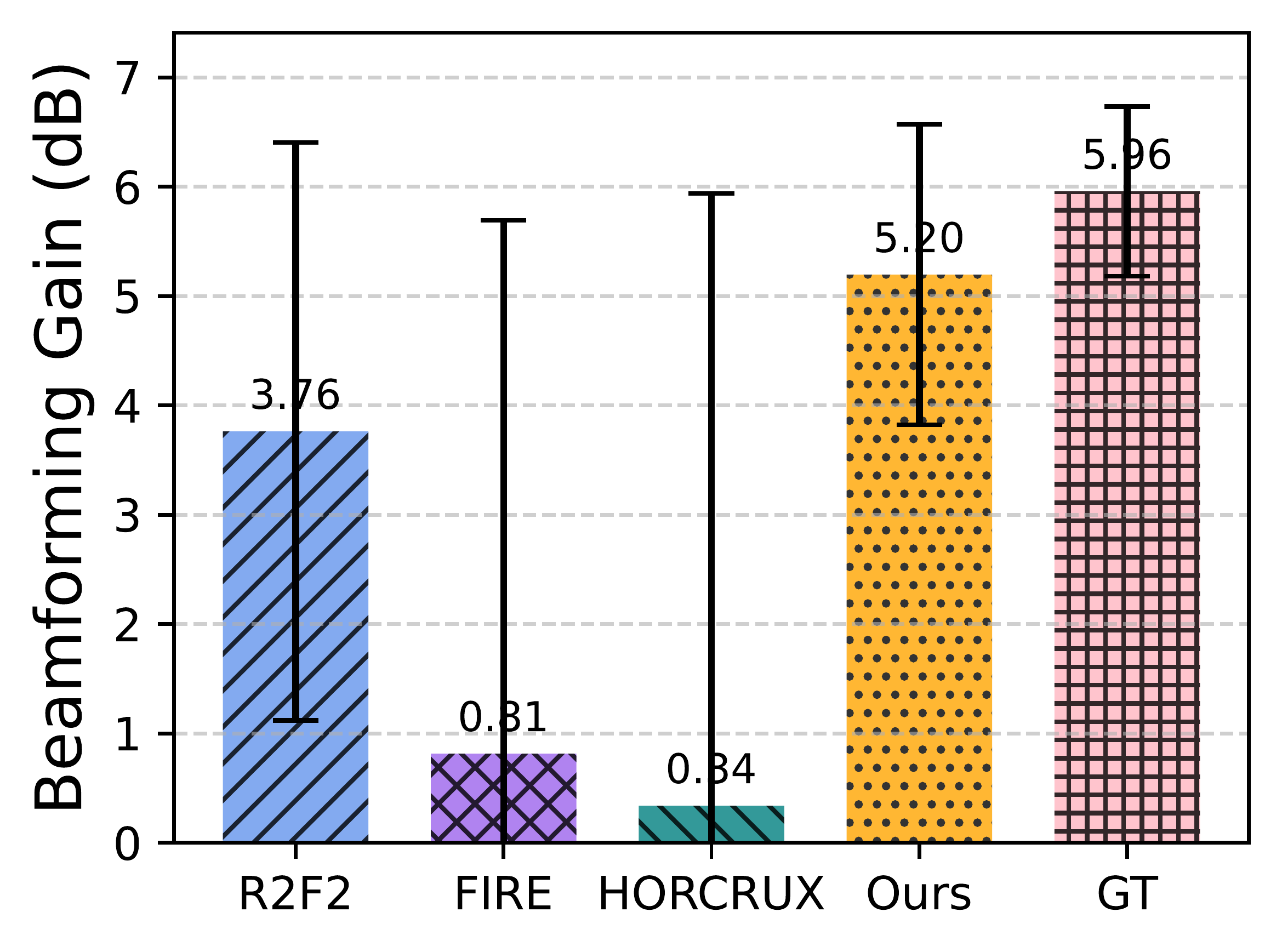}}
	\subfloat[Speed–Accuracy Trade-off.]{\includegraphics[width=0.5\linewidth]{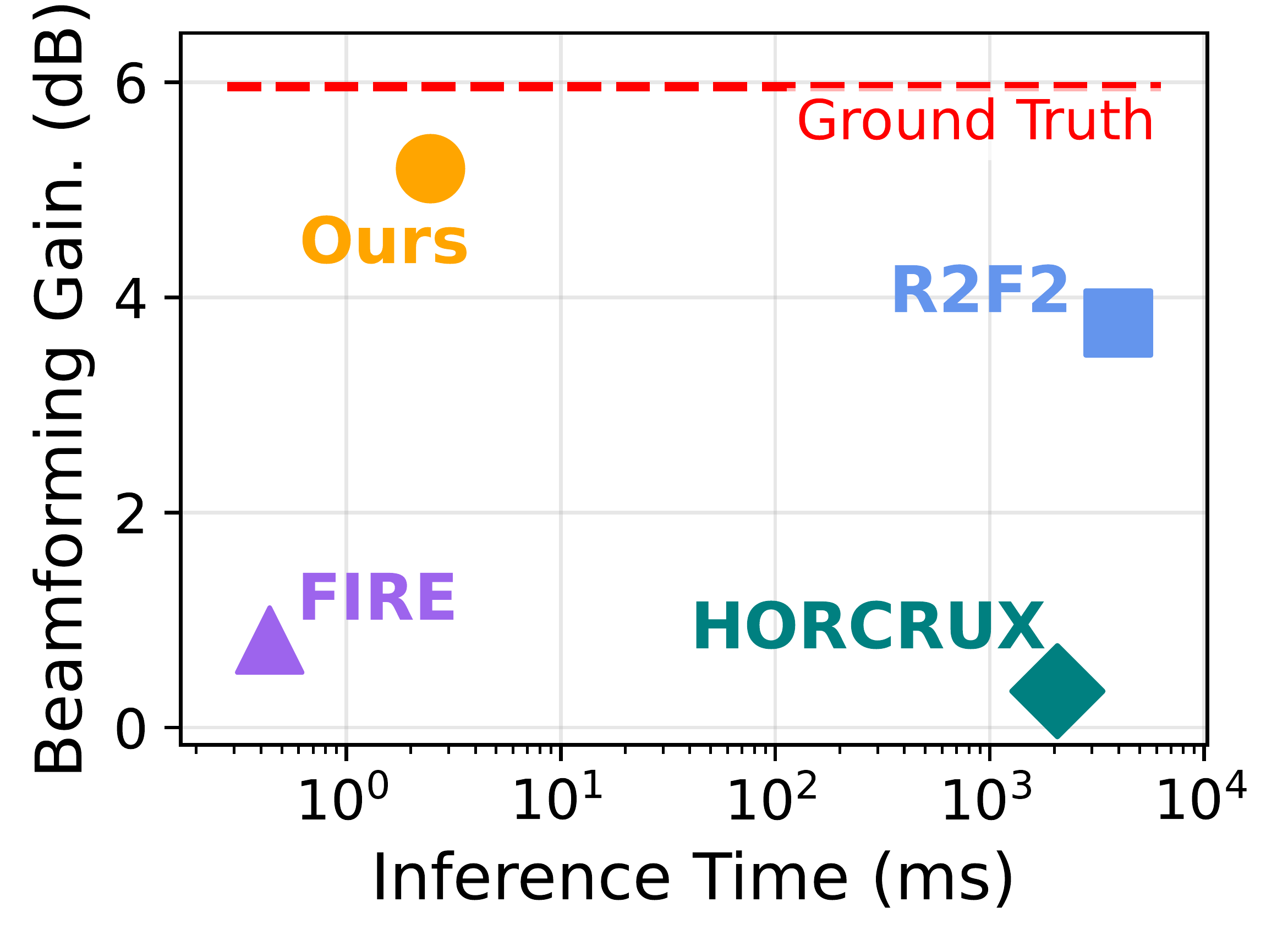}}
	\vspace{-1em}
	\caption{
		Performance comparison against baselines.
	}
	\label{fig:performance}
	\vspace{-1em}
\end{figure}

Fig.~\ref{fig:performance} compares the downstream beamforming performance and inference efficiency of the proposed method against representative baselines.
Fig.~\ref{fig:performance}(a) reports the beamforming gain across all methods in real-world measurements. 
\sys{} achieves a mean value of $5.20$~dB, outperforming R2F2 ($3.76$~dB), FIRE ($0.81$~dB), and HORCRUX ($0.34$~dB) 
by $+1.44$~dB, $+4.86$~dB, and $+4.39$~dB, respectively, and approaching within $0.76$~dB of the ground-truth (GT). The error bars in Fig.~\ref{fig:performance}(a) further show that our method provides consistently high beamforming gains across test cases. Although R2F2 can occasionally produce competitive estimates, its large variation suggests limited robustness under the evaluated conditions. FIRE and HORCRUX, on the other hand, fail to provide reliable beamforming improvement. 
Fig.~\ref{fig:performance}(b) further reveals the speed-accuracy trade-off across methods. GUIDE achieves the best of both regimes: it approaches the ground-truth beamforming gain while completing inference in under $4$~ms, over $1610\times$ faster than R2F2. Therefore, the proposed method provides a favorable balance between accuracy and efficiency, enabling a real-time AI-RAN capability for accurate channel inference and low-latency RAN optimization.

\section{Conclusion}

\sys{} achieves accurate and environment-generalizable cross-band channel prediction through a physics-guided deep unfolding design. When deployed directly from simulation to real-world measurements without retraining, \sys{} attains near-ground-truth beamforming gain at real-time inference speed and substantially outperforms all baselines. These findings suggest that \sys{} can serve as a practical and deployable channel-inference component for real-time AI-RAN.

\section{Acknowledgments}
The research work described in this paper was conducted in the JC STEM Lab of Advanced Wireless Networks for Mission-Critical Automation and Intelligence funded by The Hong Kong Jockey Club Charities Trust. The work of Ruiqi Kong is supported in part by Hong Kong ITC under Project PiH/277/25GS. The work of He Chen is supported in part by Hong Kong RGC under Project 24210524. 

\bibliographystyle{ACM-Reference-Format}
\bibliography{references}

	
	
	
	

    

\end{document}